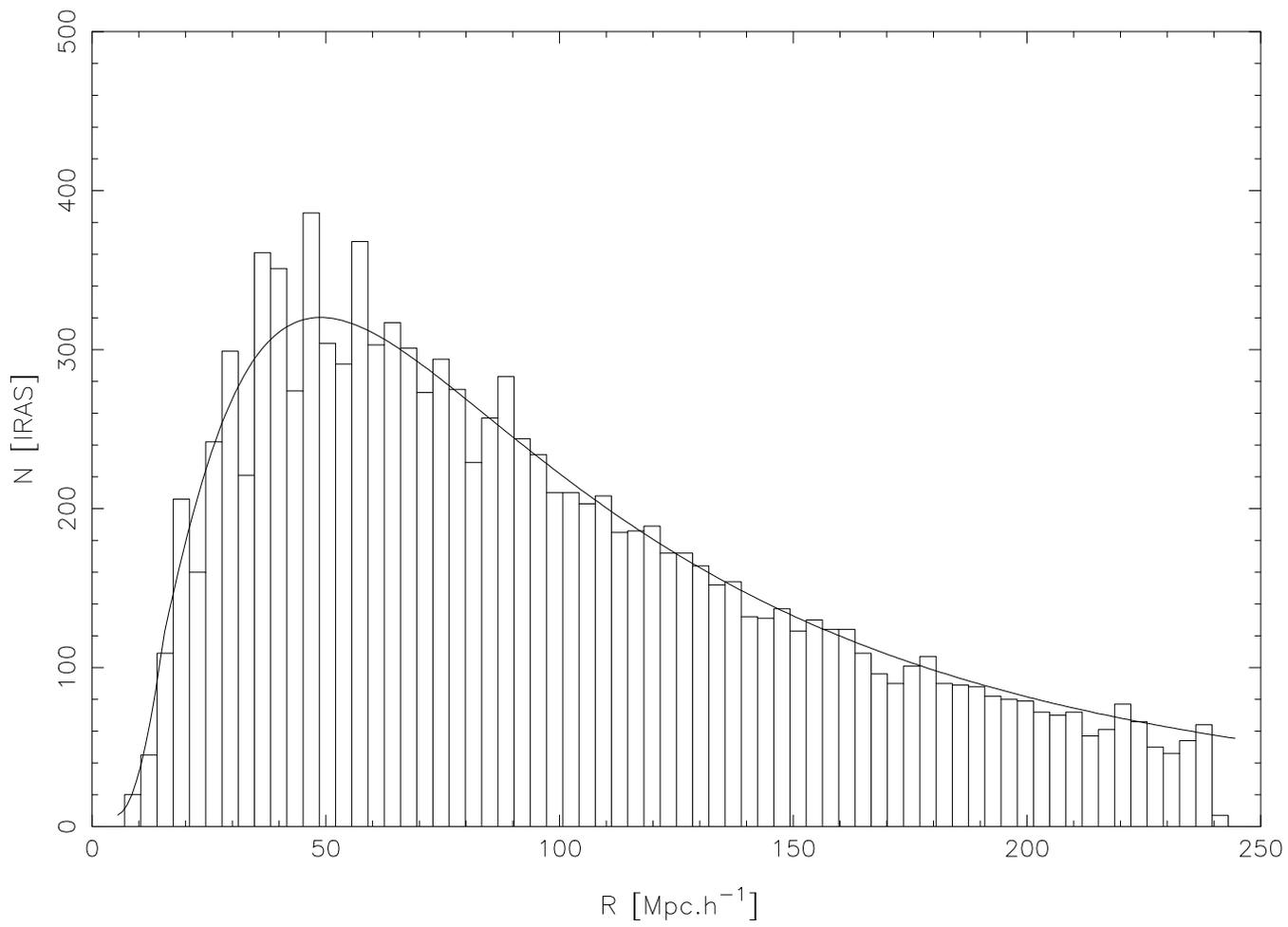

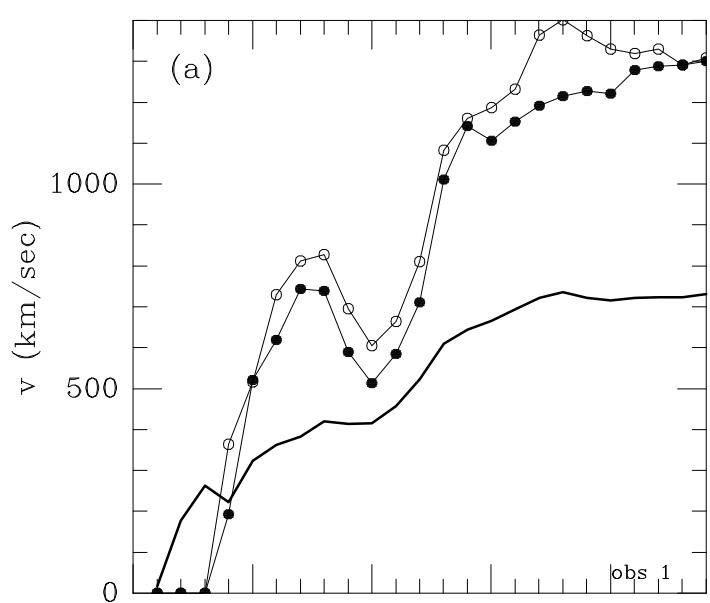
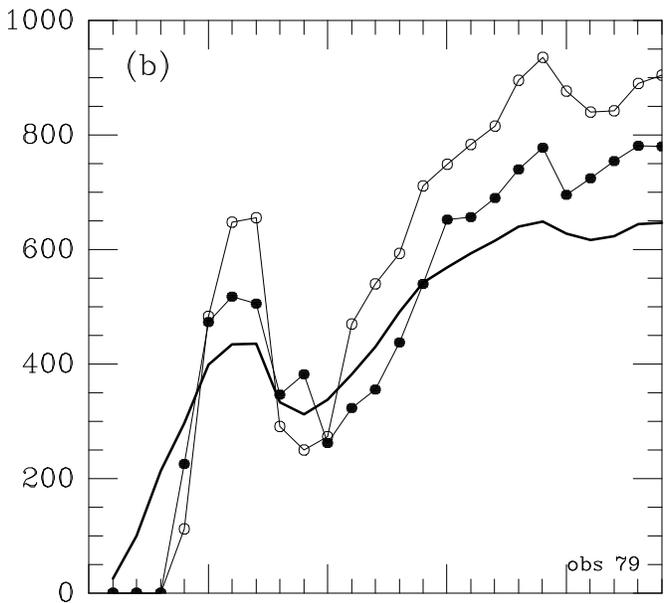
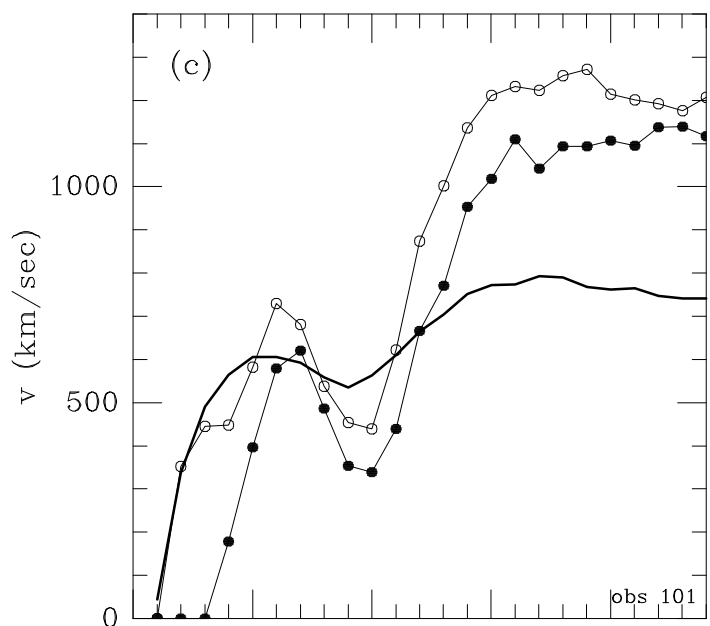
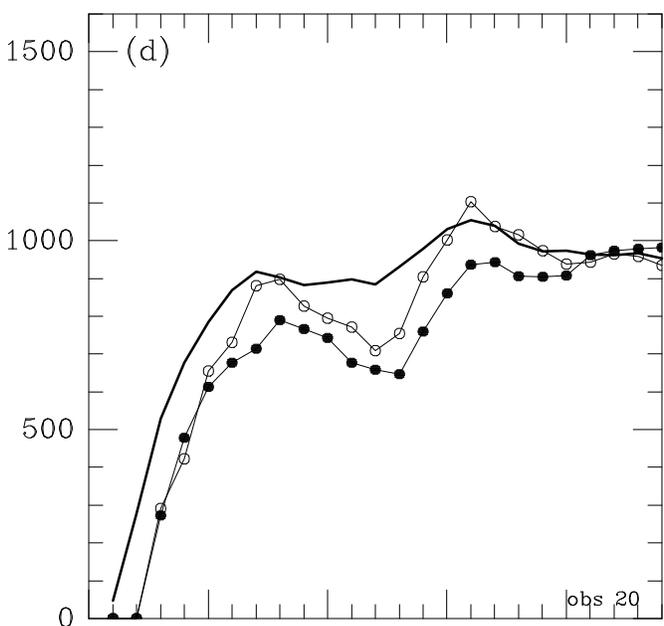
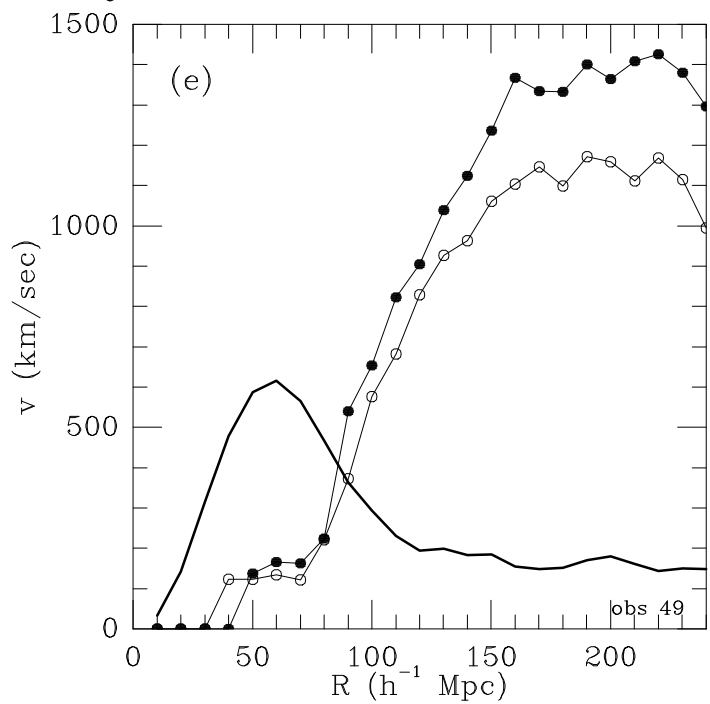
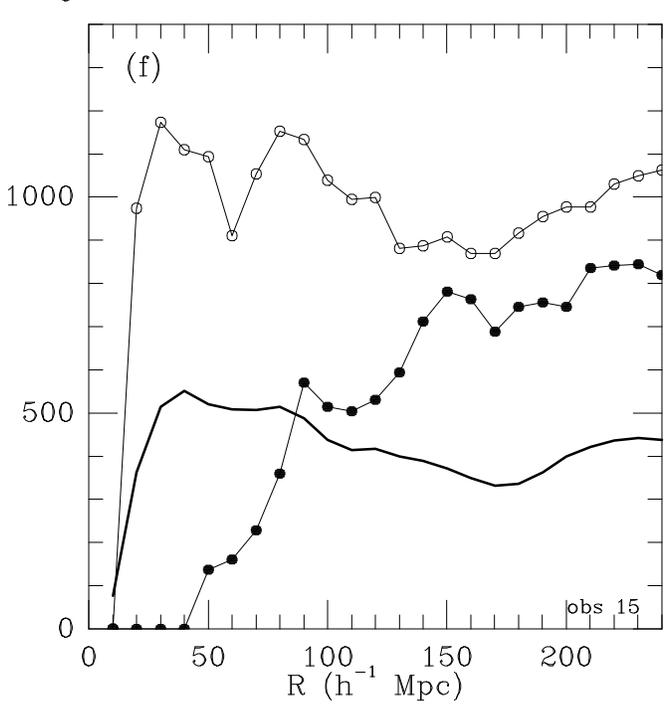

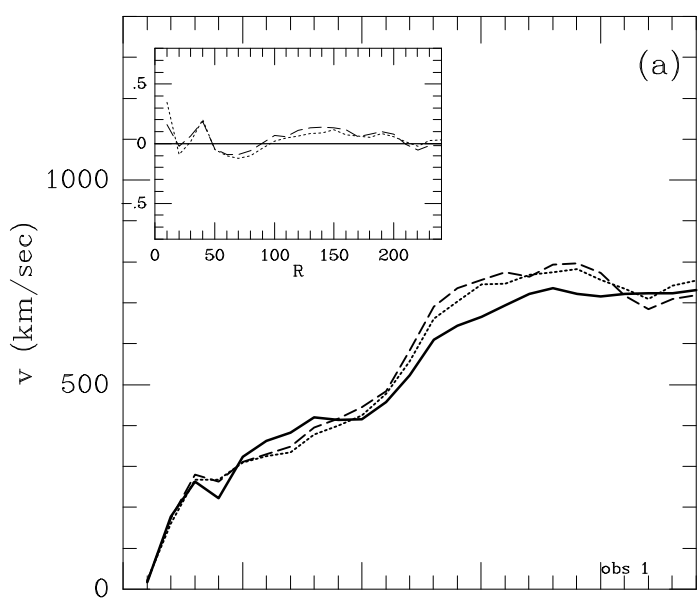
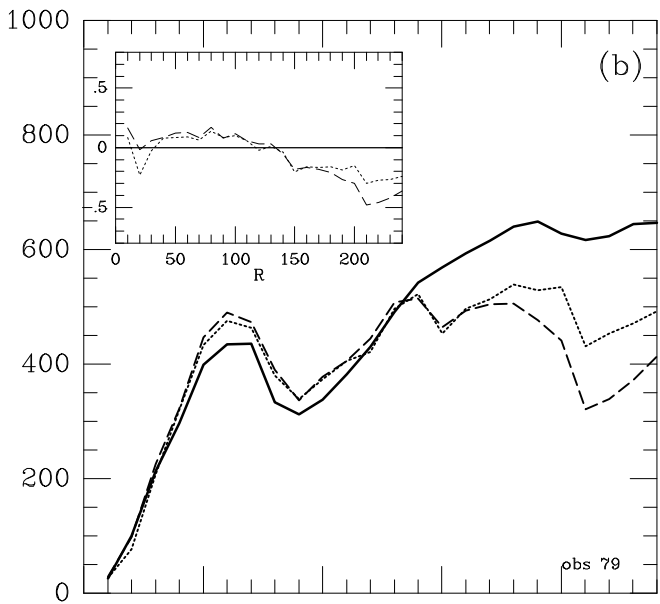
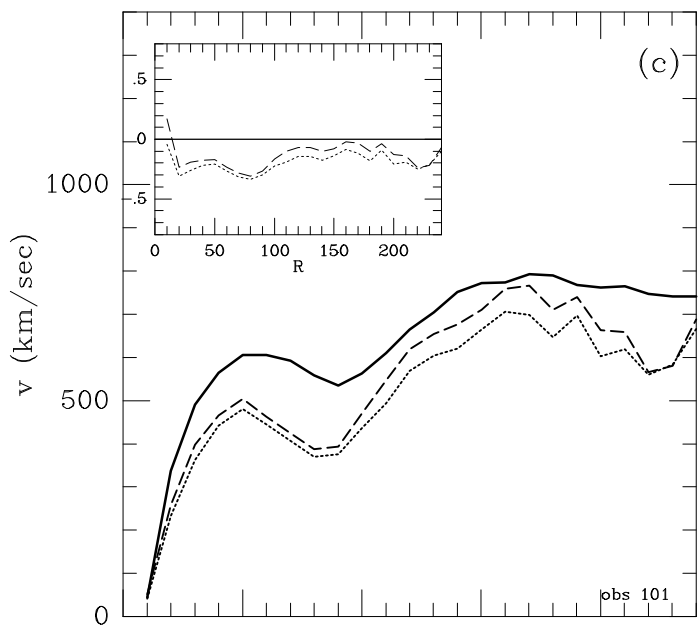
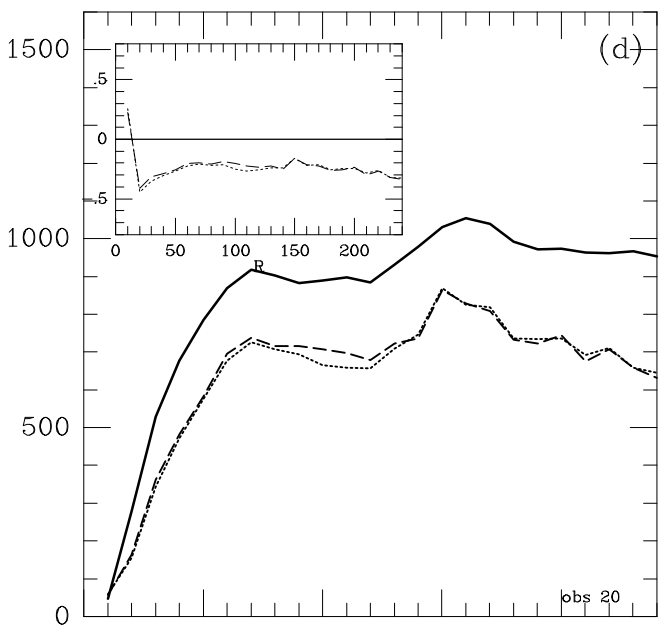
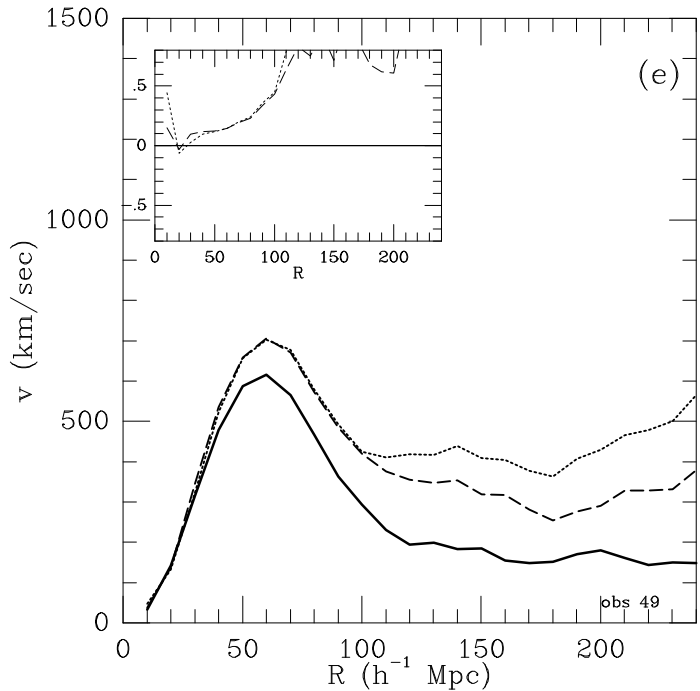
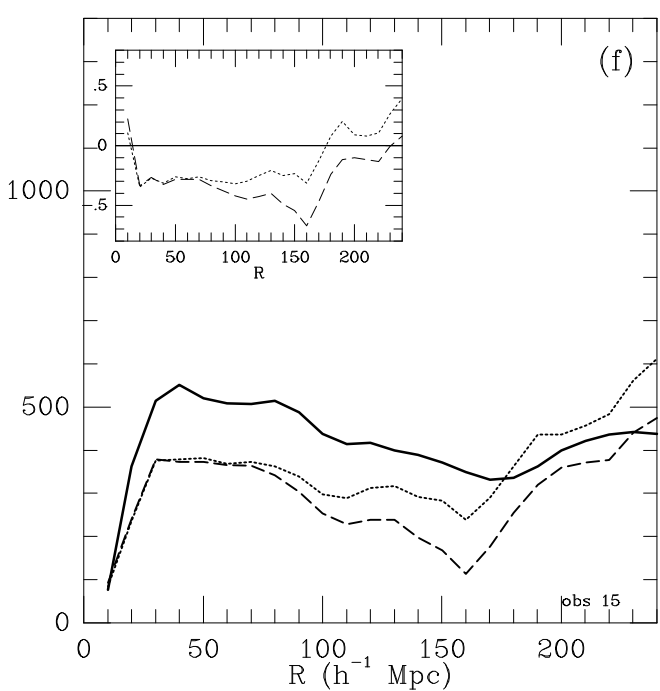

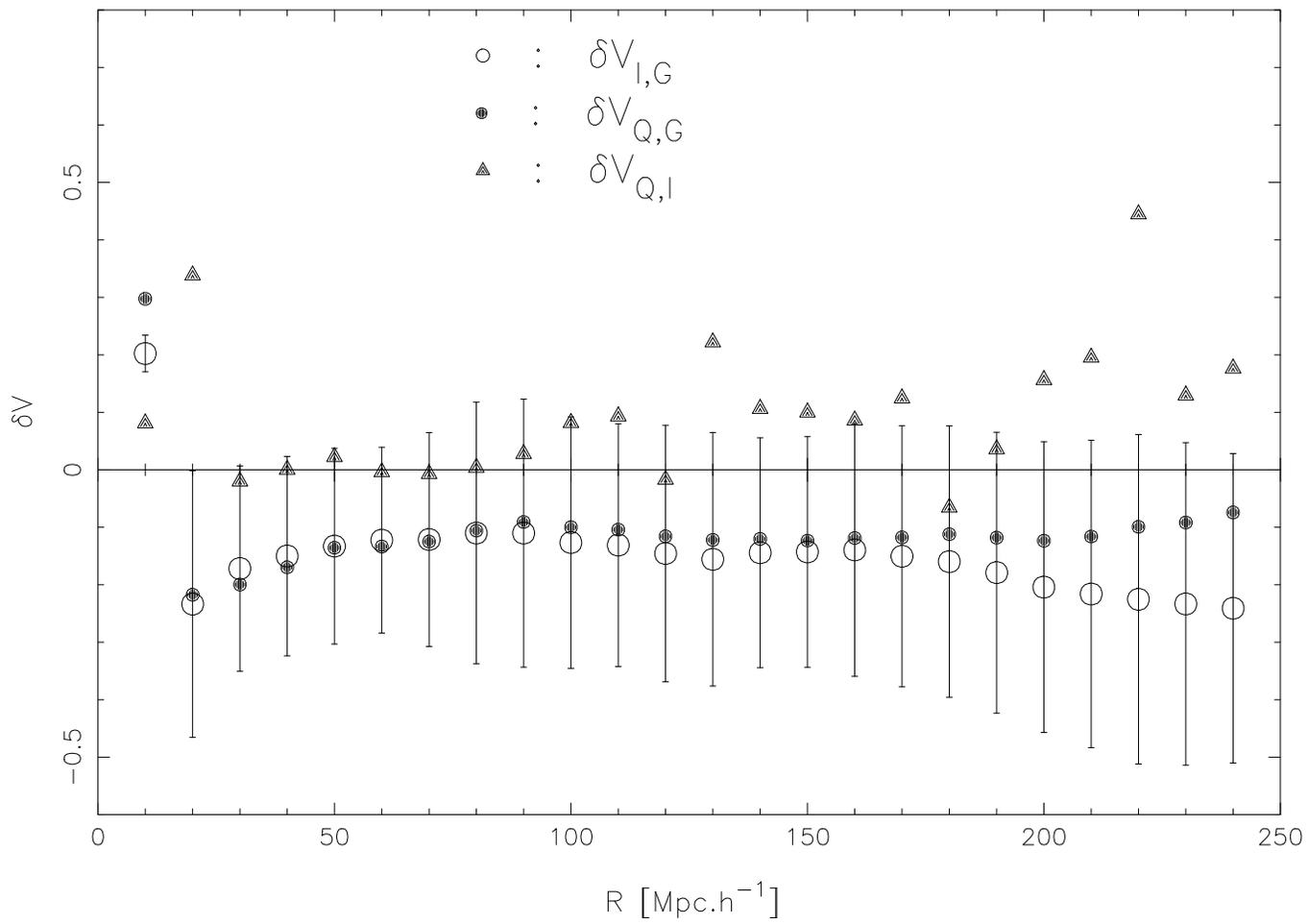

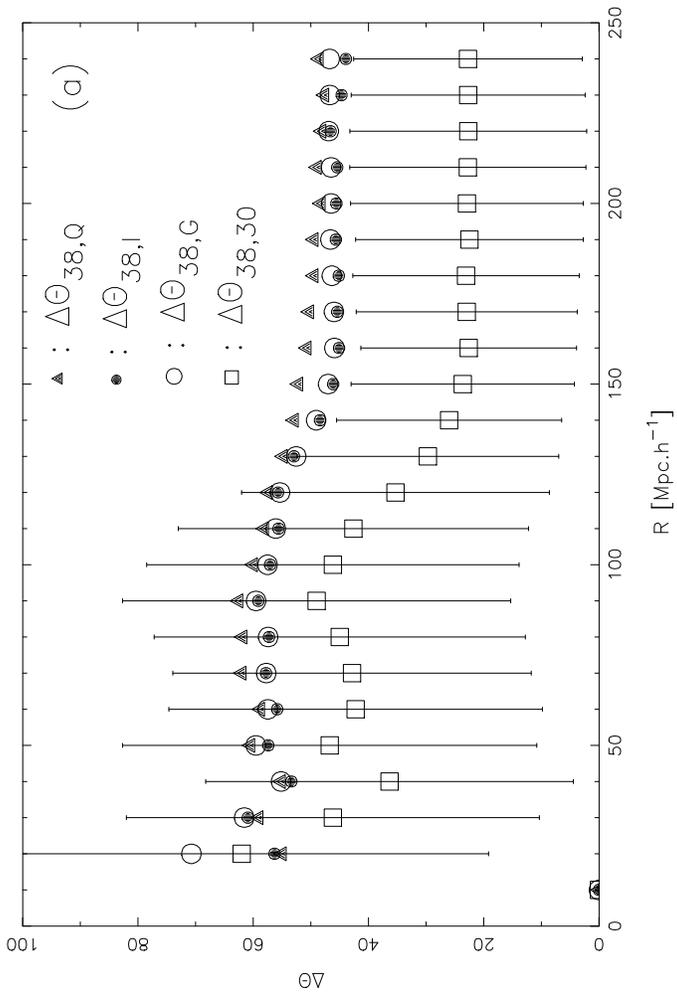
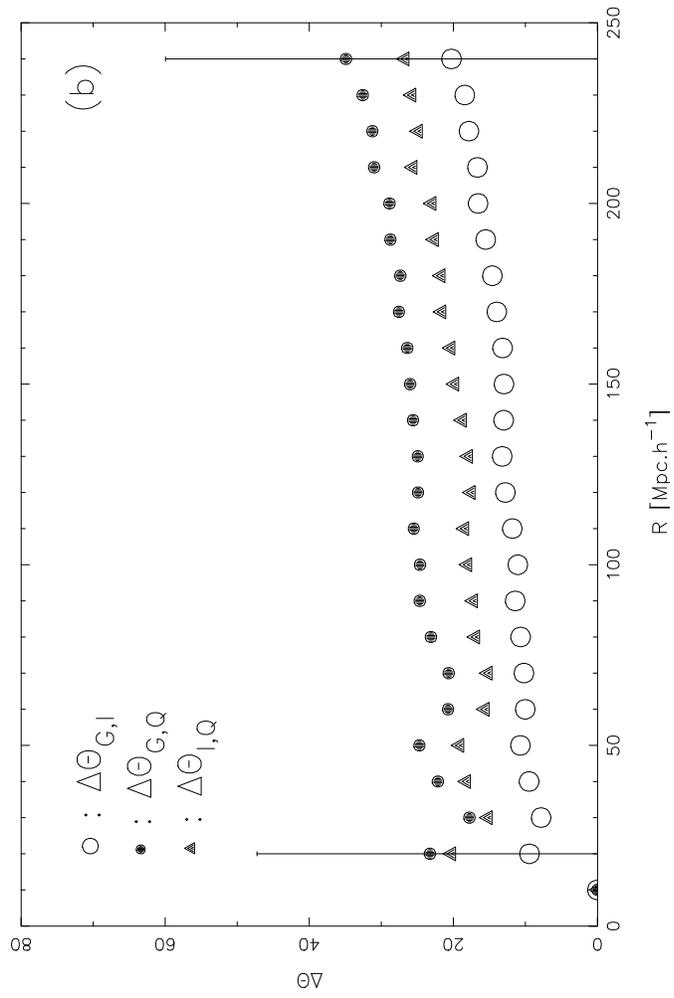

# Sampling Effects on Cosmological Dipoles

V. Kolokotronis[1], M. Plionis[2,3], P. Coles[1], S. Borgani[2,4], L. Moscardini[5]

[1] *Astronomy Unit, School of Mathematical Sciences, Queen Mary and Westfield College, Mile End Road, London E1 4NS, UK*
[2] *SISSA – International School for Advanced Studies, via Beirut 2–4, I–34013 Trieste, Italy*
[3] *National Observatory of Athens, Lofos Nimfon, Thesio, 18110 Athens, Greece*
[4] *INFN Sezione di Perugia, c/o Dipartimento di Fisica dell'Università, via A. Pascoli, I–06100 Perugia, Italy*
[5] *Dipartimento di Astronomia, Università di Padova, vicolo dell'Osservatorio 5, I–35122 Padova, Italy*

13 November 1995

**ABSTRACT**
We use numerical simulations to investigate the behaviour of the dipole moment of the spatial distribution of different kinds of mass tracers. We select density peaks of the simulated matter distribution with mean separations of 38 and 30 $h^{-1}$ Mpc to represent two samples of rich clusters of galaxies, i.e. $R \geq 0$ Abell clusters and APM clusters respectively. We also extract, from the same simulations, samples selected to mimic the full 3D galaxy distribution of IRAS galaxies, and the flux–limited IRAS and QDOT galaxy samples. We compare the dipole moments of these "galaxy" and "cluster" samples in order to assess the effects of sampling uncertainties and shot–noise on the relationship between the "true" underlying galaxy dipole and the dipoles obtained for clusters and for the flux–limited galaxy samples. The results of this analysis demonstrate that the dipoles of both the IRAS and QDOT-like catalogues should trace the full 3D dipole shape fairly accurately, with the loss however of about 15–20% of the total 3D dipole amplitude. Furthermore, using a simple argument based on linear perturbation theory, on the linear biasing assumption and on the amplitude of the cluster dipole relative to that of galaxies, we can estimate their relative biasing factors quite accurately and in agreement with results obtained by other methods.

## 1 INTRODUCTION

The issue we investigate in this paper is the extent to which different kinds of cosmic objects trace the underlying mass fluctuations. We want to see how well the large–scale density field is traced by objects selected according to different criteria. In particular we will consider IRAS/QDOT galaxies and rich galaxy clusters, as tracers of the density field.

It is usually assumed that number fluctuations of some kind of cosmic objects are related to fluctuations in mass $M$ by the linear biasing factor $b$:

$$\left(\frac{\delta N}{N}\right) = b \left(\frac{\delta M}{M}\right) . \tag{1}$$

Although there are some motivations for the presence of a form of statistical bias relating, for example, cell–count variances to mean–square matter fluctuations through a constant $b^2$ (e.g. Kaiser 1984; Coles 1993), it should be mentioned at the outset that the motivation for assuming a bias of the linear form (1) at each point is weak and this assumption may well turn out to be false. If velocity information is available and the

linear biasing assumption is adopted, then it is useful to work with the parameter

$$\beta \equiv f(\Omega_\circ)/b \simeq \Omega_\circ^{0.6}/b, \tag{2}$$

where the function $f$ is related to the rate of mass fluctuation growth in linear perturbation theory and $\Omega_\circ$ is the present–day value of the cosmological density parameter. The method used to determine the $\beta$–parameter, on which we shall concentrate here, is based on the dipole of the 3D distribution of extragalactic objects, its convergence and alignment properties. The 3D dipole moment provides an estimate of the gravitational acceleration acting on the Local Group (LG) and, in linear gravitational instability theory, the peculiar velocity vector, $\mathbf{u}(\mathbf{r})$, is proportional to, and aligned with, the peculiar acceleration vector, $\mathbf{g}(\mathbf{r})$:

$$\mathbf{u}(\mathbf{r}) = \frac{2}{3} \frac{f(\Omega_\circ) \, \mathbf{g}(\mathbf{r})}{H_\circ \Omega_\circ} = \frac{\beta}{4\pi} \int \delta(\mathbf{r}) \frac{\mathbf{r}}{|\mathbf{r}|^3} d\mathbf{r} . \tag{3}$$

Therefore, an estimate of $\beta$ can be obtained by comparing the observed LG velocity relative to the Cosmic Microwave Background (CMB) and the net gravitational force acting on it, as inferred from the observed distri-bution of cosmic objects. For this method to work, the inferred acceleration vector must be aligned with the LG velocity vector. But while this is a necessary condition for eq.(3) to apply, it is not sufficient: one must also be sure that the sample of mass tracers is sufficiently deep that it does not miss any contribution from distant (but possibly very large) density fluctuations. In other words, the dipole vector of the distribution must converge to its global value within the effective depth of the sample.

Such analyses have been performed by many authors, using different populations of extragalactic objects: optical galaxies (Lahav 1987; Plionis 1988, 1989; Lahav, Rowan–Robinson & Lynden–Bell 1988; Lynden–Bell, Lahav & Burstein 1989; Hudson 1993), IRAS galaxies (Meiksin & Davis 1986; Yahil, Walker & Rowan–Robinson 1986; Villumsen & Strauss 1987; Strauss & Davis 1988; Yahil 1988; Rowan–Robinson et al. 1990; Rowan–Robinson et al. 1991; Strauss et al. 1992; Plionis, Coles & Catelan 1993), X–ray active galactic nuclei (Miyaji & Boldt 1990), X–ray clusters (Harmon, Lahav & Meurs 1987; Lahav et al. 1989), and Abell/ACO clusters (Scaramella, Vettolani & Zamorani 1991; Plionis & Valdarnini 1991; Scaramella, Vettolani & Zamorani 1994; Tini Brunozzi et al. 1995; Branchini & Plionis 1996). The wide variety of tracers provides different values of $\beta$ which can nevertheless be reconciled with a value of $\Omega_\circ = 1$ by invoking different levels of biasing with respect to the background matter distribution. The only exception to this is the cluster dipole (Scaramella et al. 1991; Plionis & Valdarnini 1991), which, taken at face value, seems to support a value of $\Omega < 1$. Recently, however, it has been shown that redshift space distortions could be responsible for this result (Branchini & Plionis 1996): taking such effects into account, a value of $\Omega_\circ$ consistent with unity is provided also by the cluster dipole (see also Plionis 1995).

It is the purpose of this paper to investigate the way in which the dipole moments of different classes of extragalactic objects relate to that of the underlying galaxy distribution, and how sampling effects alter the convergence and alignment properties of such dipoles. In particular, we wish to address the following questions:

- Do long–range contributions to the cluster dipole (as in the Abell/ACO sample) imply similar contributions to the underlying galaxy dipole?
- Does the sparse sampling of IRAS/QDOT galaxies seriously affect the ability of such samples to trace the full 3D galaxy dipole, especially when the latter has contributions from large scales?
- What can be inferred from the observed dipole convergence and alignment properties of different mass–tracers about their relative bias parameters?

To this end we calculate the dipole vector for various simulated samples constructed in order to mimic the flux–limited IRAS/QDOT galaxy and rich galaxy cluster samples. We compare these quantities with the dipole of the full simulated IRAS–like galaxy field in order to assess the accuracy with which the samples reflect the properties of the underlying "parent" distribution. The simulations we use are based on an optimised version of the Truncated Zel'dovich Approximation (TZA).

Although this method has been demonstrated to be extremely reliable to simulate the large–scale distribution of galaxy clusters (cf. Borgani et al. 1995), it cannot be used to predict galaxy positions with any precision. However, this does not represent a limitation for the following analysis. In fact, the purpose of this paper is not to measure galaxy correlation properties, which would require a precise knowledge of galaxy positions, but instead to seek the effects of using different mass tracers to study the dynamical origin of the LG motion. It is sufficient, therefore, for us to use simulations such as these for which the large–scale matter distribution is accurately represented, even if the distribution of mass on the scale of individual galaxies is not resolved.

The layout of this paper is as follows. In Section 2 we set up the problem to be investigated and the methods we will use. In Section 3 we discuss the error analysis and in Section 4 we present our results. Finally, our main conclusions are presented in Section 5.

## 2  SETTING UP THE PROBLEM AND THE METHODOLOGY

In this paper we will deal with estimates of the peculiar gravitational acceleration induced on the positions of suitable observers, identified in a cosmological simulation, by the distribution of matter surrounding them. Such estimates will be given by the dipole moment of the spatial distribution, around the observer, of some kind of mass–tracer. In order to answer the questions posed in the introduction we need to introduce *(a)* the moment–based method to estimate the acceleration vector for a model simulated universe; *(b)* the procedure to identify suitable observers; *(c)* the method to extract realistic IRAS, QDOT and cluster samples for each of these observers.

Note that we are not interested in studying the effects of incomplete sky coverage, Galactic absorption, Virgocentric infall corrections and redshift space distortions. For such a study and its implications for dipole analysis, see Tini Brunozzi et al. (1995) and references therein.

### 2.1  The formalism

Given a set of objects at positions $\mathbf{r}_i$, the peculiar gravitational acceleration, acting on the observer at the origin of the coordinate system, is estimated from the dipole moment:

$$\mathbf{D} = \sum_{i=1}^{N} w_i \hat{\mathbf{r}}_i \ , \qquad (4)$$

where $\hat{\mathbf{r}}_i$ is the unit vector pointing at the position of each object and $N$ is the total number of such objects within the distance considered. The weights, $w_i$, should be of the form $w_i \propto r_i^{-2}$ (since both flux and gravity fall as $r^{-2}$).

Note that the dipole, as a function of distance from the observer, keeps rising till isotropy is reached, after which it flattens out and converges to its final value. Similarly, the monopole moment $M$ is defined as

$$M = \sum_{i=1}^{N} w_i . \tag{5}$$

According to its definition, the monopole term is related to the number density of objects as a function of radial distance; in the continuous case we have that $M(r) = 4\pi\langle n\rangle r$. Therefore, it is evident that we can define an estimator for the average density $\langle n\rangle$ within a distance $R$ as

$$\bar{n}_1 = \frac{M(R)}{4\pi R}, \tag{6}$$

where $R$ is taken to be sufficiently large (at least of the order of $R_{conv}$, the dipole convergence scale), so that the effects of local density fluctuations cancel out in $M$. Another estimator of the mean object space density can be given by the inverse cube of their mean separation:

$$\bar{n}_2 = \langle d\rangle^{-3} . \tag{7}$$

In the most common cases of magnitude– or flux–limited extragalactic object catalogues one has to take into account the effects of the consequent undersampling of the density field especially at large distances, where the radial selection functions rapidly decline. Assuming that the unobserved galaxies are spatially correlated with those included in the catalogue, the usual procedure to correct for the missing population is to weight each observed galaxy at a distance $r$ by $1/\phi(r)$, which is the reciprocal of the portion of the luminosity function that cannot be sampled at that distance due to the flux or magnitude limit of the catalogue. Therefore the weight in eq.(4) and eq.(5) should become $w_i \simeq \phi^{-1}(r_i)\, r_i^{-2}$, where

$$\phi(r_i) = \int_{L_{min}(r_i)}^{L_{max}} \Phi(L)dL \tag{8}$$

is the selection function. The lower limit of this integral, determined by the flux limit, is the minimum luminosity that an extragalactic object can have in order to be visible at a distance $r$, while $L_{max}$ is the maximum luminosity of such an object. Note that the inverse selection function weight grows with increasing $r$ and this acts as a compensating effect for the fact that poor sampling occurs at large redshifts.

Finally, using linear perturbation theory (eq.3) and the above definition of the dipole we can relate the observed peculiar velocity of an observer, $u$, with that predicted by its gravitational acceleration, $v$, for as long as the two vectors are well aligned:

$$u(r) = \frac{H_\circ\beta}{4\pi\bar{n}} \sum_{i=1}^{N} w_i \, \frac{\mathbf{r}_i}{|\mathbf{r}_i|^3} = \beta\, v(r) . \tag{9}$$

This relation can in principle provide an estimate of the $\beta$–parameter.

In the following analysis, we will explicitly assume that there are no appreciable density fluctuations at distances larger than our sample limiting depth ($R_{max} = 240\ h^{-1}$ Mpc). Therefore, the density fluctuations causing the motion of the observers is assumed to be well sampled by the distribution of objects lying within $R_{max}$, which is also consistent with the observed convergence of the cluster dipole at $\simeq 170\ h^{-1}$ Mpc.

## 2.2 The Simulations

The simulations of the large–scale matter distribution that we use in this work are generated by using an optimized version of the Truncated Zel'dovich Approximation (TZA) described in detail in Borgani et al. (1995). Here we only briefly remind the reader of its main ideas. Particles are moved from their initial (Lagrangian) positions to their final (Eulerian) coordinates by the usual Zel'dovich mapping, $\mathbf{x}(\mathbf{q}, t) = \mathbf{q} + b(t)\nabla\psi(\mathbf{q})$, where $b(t)$ is the perturbation linear growth factor and $\psi(\mathbf{q})$ is the velocity potential, which is related to the initial density fluctuation field by the Poisson equation $\nabla^2\psi(\mathbf{q}) = -\delta(\mathbf{q})/a$. According to this prescription, particles move in straight line trajectories with velocity $\mathbf{v} = \dot{b}(t)\psi(\mathbf{q})$. This gives a reasonably accurate account of the particle motions until "shell–crossing" occurs, i.e. until the mapping between initial and final positions becomes degenerate. In practice, we reduce the amount of shell–crossing by suitably filtering the short wavelength modes, by using a Gaussian filter with radius chosen in such a way that $N_s = 1.1$ for the average number of streams at each Eulerian point. Although this filtering procedure restricts the amount of non–linearity the TZA can cope with, nevertheless the optimized version of this method is well–suited to the problem of simulating the distribution of rich clusters, in which the smoothing scale required is similar to the "catchment area" of a protocluster.

Velocity and density fields are reconstructed on $256^3$ grid points, using as many particles, for a $L = 480\ h^{-1}$ Mpc simulation box. This size is a compromise between the need of using large samples to ensure dipole convergence and the need for good grid resolution to identify mass tracers accurately. As usual, the box has periodic boundaries.

The following analysis is based on a realization of a Cold+Hot dark matter (CHDM) model with $\Omega_\circ = 1$, $\Omega_{hot} = 0.3$ and $\Omega_b = 0.1$ for the contributions of hot and baryonic component respectively, $H_\circ = 50$ km s$^{-1}$ Mpc$^{-1}$ for the Hubble constant and $\sigma_8 = 0.78$ for the r.m.s. fluctuation amplitude within a top–hat sphere of $8\ h^{-1}$ Mpc radius (see Klypin, Nolthenius & Primack 1995, and references therein, for the cosmological relevance of this model). This spectrum normalization corresponds to $Q_{rms-PS} = 20\ \mu K$ for the quadrupole of the CMB temperature anisotropy from the two–year *COBE* data (Górski, Stompor & Banday 1995).

Although different dark matter models can easily be simulated with our TZA method, our main interest here is not to use the dipole analysis as a test for such models (see Tini Brunozzi et al. 1995, for such an approach). This is the reason why we concentrate here only on the CHDM model, which we showed anyway to provide an excellent description of the cluster correlation properties (Plionis et al. 1995; Borgani et al. 1995). Furthermore, although many realizations of the same model can be quickly generated, we consider here a single model realization since we verified that any "cosmic variance" effect on dipoles is dominated by the observer–to–observer scatter, rather than by differences between different such big simulation volumes.

### 2.3    Choosing suitable observers

We now proceed in identifying in our simulation suitable, for our purpose, observers. We would ideally prefer to choose observers that are LG–like. However, as we shall see, we have found very few such observers and thus we will define observers with only some of the LG characteristics. The criteria for selecting LG–like "observers" are:

(i) Peculiar velocity of $627 \pm 44$ km s$^{-1}$ (error corresponding to $2\sigma$ uncertainties; Kogut et al. 1993) for a top–hat sphere of $7.5\,h^{-1}$Mpc radius centred on the observer.

(ii) Density contrast of $-0.2 \leq \delta_{LG} \leq 1$ within the same sphere.

(iii) Dipole shape for the surrounding cluster distribution similar to the real Abell/ACO cluster dipole (Scaramella et al. 1991; Plionis & Valdarnini 1991; Branchini & Plionis 1996), i.e. a two–step increase of the amplitude and final dipole convergence at $\sim 140 - 180$ $h^{-1}$ Mpc.

The first two criteria are the usual ones invoked in numerical simulation studies (cf. Moscardini et al. 1995; Strauss et al. 1995; Tini Brunozzi et al. 1995). The last criterion is imposed in order to reproduce the observed Abell/ACO galaxy cluster dipole, since (as discussed in the introduction) we want to investigate *(a)* whether the particular cluster dipole shape implies that a similar dipole shape should be observed in the underlying galaxy distribution and *(b)* whether flux–limited IRAS subsamples can *see* such dipole contributions from large distances ($\gtrsim 120\,h^{-1}$ Mpc).

Out of the original $N_{obs}$=20000 random observers selected in our simulation we found only 7 fulfilling all three criteria. Since this number is extremely small for a statistical study and since the questions we want to answer are strictly related to the last criterion, we relaxed the first two, obtaining in total 106 observers. Our analysis will be based on this set of observers.

### 2.4    Mass–tracers selection

In what follows we present the method of extracting realistic distributions of clusters and galaxies from our simulation. Note that the distribution of simulated clusters will closely resemble the corresponding distribution of Abell/ACO clusters (cf. Borgani et al. 1995). However, we will identify galaxies as simulation particles having the observed large–scale characteristics without implying that their detailed distribution and clustering properties should resemble those of real galaxies; our simulation is just not well suited for such a task.

[tp]

**Table 1.** Parameters for the Saunders et al. (1990) luminosity function.

| Parameters | Values | Units |
|---|---|---|
| $C$ | $0.026 \pm 0.008$ | $h^3$ Mpc$^{-3}$ |
| $\alpha$ | $1.09 \pm 0.12$ | – |
| $\sigma$ | $0.724 \pm 0.031$ | – |
| $L_*$ | $10^{(8.47 \pm 0.23)}$ | $h^{-2} L_\odot$ |

#### 2.4.1    Simulated Galaxy Clusters ($\equiv C$)

We identify clusters as local peaks of the evolved density field, after reconstructing it at the grid positions. We select two populations of clusters; one with a mean intercluster separation of $\langle d_C \rangle = 38\ h^{-1}$ Mpc ($C38$ hereafter), which corresponds to the $R \geq 0$ Abell/ACO sample, and one with $\langle d_C \rangle = 30\ h^{-1}$ Mpc ($C30$ hereafter), corresponding to the APM sample (Dalton et al. 1994). The total number of clusters within the box (i.e. the highest local density maxima on the grid) is given by $N_{C,i} = (L/\langle d_C \rangle)^3$. Thus we have $N_{C,1} = 2015$ and $N_{C,2} = 4096$ for the two populations, corresponding to densities of $\bar{n} \approx 1.82 \times 10^{-5}\ h^3$ Mpc$^{-3}$ and $\bar{n} \approx 3.7 \times 10^{-5}\ h^3$ Mpc$^{-3}$, respectively.

#### 2.4.2    Simulated IRAS Galaxies ($\equiv G$)

In order to simulate the IRAS galaxy sample we need to reproduce its observational flux limit. This, in turn, requires galaxy luminosities to be drawn from the appropriate luminosity function,

$$\Phi(L) = C \left( \frac{L}{L_*} \right)^{1-\alpha} \exp \left[ -\frac{1}{2\sigma^2} \log_{10}^2 \left( 1 + \frac{L}{L_*} \right) \right] \quad (10)$$

(Saunders et al. 1990), the parameters of which are given in Table 1. We have used only the central values of these parameters in this analysis, but we do not expect the results to be very sensitive to changes within the known error limits. We should also mention at this point that there is possibly a significant effect due to evolution of the luminosity function (cf. Saunders et al. 1990). We shall not take this into consideration here but, if it were present over the redshift range we are considering here, it could have a big effect on the convergence properties of the IRAS/QDOT dipole (Plionis et al. 1993).

To estimate the expected number density, $\langle n_G \rangle$, of the underlying distribution of IRAS–like galaxies we integrate $\Phi(L)$ over the luminosity range that we want to represent:

$$\langle n_G \rangle = \int_{L_{\min}}^{L_{\max}} \Phi(L) \mathrm{d}L \ , \quad (11)$$

where we take $L_{\min}$ and $L_{\max}$ to be $10^8$ and $10^{13} L_\odot$ respectively: this gives $\langle n_G \rangle \approx 0.031\ h^3$ Mpc$^{-3}$. The lower limit in $L$ is imposed by the fact that the very low luminosity galaxies are not well represented in the IRAS sample that we want to simulate (Rowan–Robinson et al. 1990, hereafter RR90). By multiplying this value by the volume of the simulation box used here, we obtain the total number of IRAS galaxies expected within the volume considered, $N_G \approx 3.5 \times 10^6$ objects. We therefore randomly select $N_G$ objects from the total of $256^3$ simulation particles which will represent the underlying distribution of IRAS–like galaxies from which flux–limited subsamples will be extracted. Therefore, the bias factor of this "galaxy" population is by definition $b_G = 1$.

### 2.4.3 Simulated flux–limited IRAS galaxies ($\equiv I$)

To simulate the IRAS flux–limited catalogue we use the following prescription. We take the number density of objects lying in a shell of width $\Delta r$ at distance $r$ to be

$$N(r) = 4\pi r^2 \langle n_G \rangle \phi(r) \Delta r, \qquad (12)$$

where $\langle n_G \rangle$ is given by eq.(11) and $\phi(r)$ is estimated from eq.(8) by using

$$L_{\min}(r) = 4\pi r^2 S_{\lim} \nu_{60} . \qquad (13)$$

Here $S_{\lim}$ corresponds to the QDOT flux limit of 0.6 Jy and $\nu_{60}$ is the frequency of the passband at $60\,\mu m$ which is used to convert fluxes to luminosities and $L_{\max} = 10^{13} L_\odot$ (RR90). 

The redshift distribution of the resulting $N_I \sim 12000$ mock IRAS galaxies together with the theoretical curve of eq.(12) is shown in Figure 1. It closely resembles the corresponding QDOT distribution (cf. Figure 2 of RR90); it has a maximum at $\sim 50\ h^{-1}$ Mpc ($z \sim 0.018$) and there is an extended tail out to the limiting depth $R_{\max}$. The above number of IRAS galaxies expresses the number of objects that can be seen above the flux limit of 0.6 Jy.

Evidently the number–density of objects, $\bar{n}$, is of major importance in the sample definition. For this reason, we have checked the various density estimators eqs. (6) and (7), which give a maximum difference of $\delta \bar{n} \approx 8\%$ and are therefore in very good agreement.

### 2.4.4 Simulated QDOT galaxies ($\equiv Q$)

Having generated IRAS–like catalogues down to a flux limit of 0.6 Jy, our next task is to build QDOT mock catalogues which constitute a random 1–in–6 subsample of the complete catalogue. In order to check the statistical reliability of the results, we repeated the random 1–in–6 selection procedure 30 times for each selected observer using different initial seeds. Clearly, the QDOT–like catalogues have $N_Q \sim 2000$ objects each.

### 2.4.5 Assigning appropriate weights

To estimate the cluster ($C$) and full IRAS ($G$) dipole we use their whole 3D distribution. Note that clusters and galaxies are not in fact point–like objects and thus an appropriate weight should really take account of their mass, i.e. $w_i \simeq M_i/r_i^2$. However, for sake of simplicity and uniformity, we will assume unit masses for all available objects. However, we verified that no serious discrepancy occurs in the dipole convergence, amplitude or alignment properties with the use of either weights.

In the case of flux–limited IRAS samples ($I \& Q$), we use the inverse selection function weights. Since it would be computationally expensive to solve the integral in

eq.(8) for all the objects in each sample, we create a list of 240 $r_i$ distance bins having a 1 $h^{-1}$ Mpc width each and evaluate the integral in eq.(8) at the centre of each distance bin and thus recover the depth dependance of the selection function. Then we compare each true object distance $r_j$ with this list of distances and assign the appropriate weight to the galaxy that corresponds to $r_i$, as long as $r_i - 1 < r_j < r_i + 1$. The algorithm thus allows us to use the previously computed and tabulated $\phi(r_i)$ instead of the true $\phi(r_j)$. Clearly, this is an approximation and it could introduce small errors. We have found in tests that the corresponding relative difference in the weights $w_i$ between the true and the above estimate of the selection function is of the order of $\sim 10^{-4}$. Note that in a perfectly uniform distribution, the overestimate/underestimate of $\phi^{-1}(r_i)$ as produced by the above scheme, would generate similar effects, which we would expect to cancel out within a sufficiently large volume. Thus the above selection function simplification should not introduce any spurious modification on the structure, convergence and alignment of the estimated dipoles.

With the above prescription to extract the different mass tracer populations and to define the appropriate weights to be used in eq.(4) and eq.(5), we proceed to calculate the dipole and monopole moments of the spatial distribution of these objects. To do so, we place each observer at the origin of the coordinate system and use the periodic boundaries of our simulations to ensure full spherical coverage.

## 3 ERROR ANALYSIS

### 3.1 Stability of dipole estimators

We estimate the dipole and monopole moments for all the datasets extracted in the previous section using 7.5 $h^{-1}$ Mpc, as the minimum distance, so as to be consistent with the definition of observer velocity and local density, and integrating up to 240 $h^{-1}$ Mpc (the simulation box limit) using bins of 10 $h^{-1}$ Mpc width. Note that in eq.(9) the normalization is based on the estimated value of the mean density of objects ($\bar{n}_1$), which is computed according to eq.(6) as the average over the last four bins, [210–240] $h^{-1}$ Mpc, to reduce the effects of statistical fluctuations in the value of $M$. We checked that estimating $\bar{n}_1$ over the [170–210] $h^{-1}$ Mpc range gives very small differences in the dipole amplitudes ($\delta D \sim 1\%$), convergence properties and alignments.

Many authors have argued how the dipole convergence could be artificially imposed by undersampling on large depths which causes also the monopole to flatten. As can be seen from eq.(6) and eq.(9), a possible early leveling of the monopole moment could cause the dipole to flatten artificially (since the dipole gains amplitude much more slowly than the monopole). The monopole of all available datasets was found to be linearly rising up to $R_{\max}$. Note however that although this is a necessary condition for the reality of the dipole convergence it is not, by any means, a sufficient condition. In a

flux–limited catalogue you can have a linearly increasing monopole merely due to the volume preserving correction factors, $\phi^{-1}(r_i)$, although the density field may be extremely sparsely traced at large distances and therefore the dipole could be underestimated at such distances. As we will see there is such evidence from the flux–limited IRAS–like samples.

## 3.2 Shot–noise estimators

Due to the sparse sampling of the full IRAS density field by the flux–limited subsamples ($I\&Q$), their moments are affected by discreteness effects (shot–noise errors), that increase with redshift because of the rapidly declining selection function. These effects introduce a variance in the dipole amplitude and direction. We use two different methods to estimate the shot–noise effects.

The first method is similar to that used by Plionis et al. (1993). For each IRAS–like sample we generate 30 random realizations in which we reshuffle the position on the sky of each galaxy while retaining its true distance. This approach is computationally very fast and preserves the exact shape of IRAS selection function. The shot–noise dipole for each resampling is

$$\mathbf{D}_{SN} = \sum_{i=1}^{N} w_i \hat{\mathbf{r}}_{SN,i}, \tag{14}$$

where $\hat{\mathbf{r}}_{SN,i}$ are the new random directions on the sky.

A different procedure to estimate the shot–noise dipole (Strauss et al. 1992; Hudson 1993; Branchini & Plionis 1996) is to write its components along the three axes, as $\sigma_x^2$, $\sigma_y^2$, $\sigma_z^2$. For example for the $z$–axis we have

$$\sigma_z^2 = \sum_{i=1}^{N} w_i^2 \left( \frac{z_i}{|r_i|^3} \right)^2, \tag{15}$$

where $\sigma_{3D} = \sqrt{\sigma_x^2 + \sigma_y^2 + \sigma_z^2}$ and the corresponding error along the line of sight (1D error) is $\sigma_{1D} = \sigma_{3D}/\sqrt{3}$. By comparing the two methods we find that $\sigma_{3D} \approx |\mathbf{D}_{SN}|$ with a maximum difference of $\sim 8\%$ for the $I$–sample and $\sim 11\%$ for the $Q$–sample.

Generally we find that $|\mathbf{D}_{SN}|$ is approximately $(10-35)\%$ of the overall $I$ and $Q$ sample dipole value while the shot–noise errors for the $Q$–samples are almost twice as large as the corresponding for the $I$–samples: this is obviously due to the much sparser sampling of the density field in the former case.

Therefore, our procedure for the error analysis can be sketched as follows.

• We will calculate shot–noise errors using the first method of eq.(14). Since it is Monte Carlo based, for each observer we will estimate the mean value of $\mathbf{D}_{SN}$ by averaging over 30 realizations of each sample. Note that the second method produces always shot–noise values within $\pm 1\sigma$ of those provided by the first method.

• Each estimated dipole (of the $I$ and $Q$ samples) will be corrected for the shot–noise effects by subtracting the 1D shot–noise dipole. Therefore we will be discussing *corrected* dipole values of the form :

$$|\mathbf{D}_{cor}| = |\mathbf{D}_{raw}| - |\mathbf{D}_{SN}|/\sqrt{3}. \tag{16}$$

• We do not apply any correction for the variance introduced by assuming a unit mass for all objects (cf. Strauss et al. 1992; Branchini & Plionis 1996).

## 4 RESULTS AND DISCUSSION

### 4.1 Dipole comparisons and Velocity fluctuations

In order to realize a quantitative comparison between the different mass tracer dipole shapes, we will perform a correlation coefficient analysis for all possible combinations of samples. This is solely based on the shape of the dipole as a function of radial depth and not on the dipole amplitudes, which in the case of galaxies vs. clusters are expected to differ significantly due to their different bias parameters. The correlation coefficient is defined as

$$c_{ij} = \frac{\sum_{i=1}^{N} (x_i - \bar{x})(y_i - \bar{y})}{\sqrt{\sum_{i=1}^{N} (x_i - \bar{x})^2} \sqrt{\sum_{i=1}^{N} (y_i - \bar{y})^2}}, \tag{17}$$

where the quantities $x_i, y_i$ represent the different mass tracer dipoles and the sum is over all (or a selected range of) the radial bins. We will present the average value, $\langle c_{i,j} \rangle$, over the ensemble of the selected 106 observers.

Furthermore, since the different galaxy dipoles should trace the same underlying fluctuations and have, by definition, $b_G = b_I = b_Q$, we will assess how well the flux–limited samples trace the full galaxy dipole by computing their relative velocity fluctuations as a function of distance, averaged again over all observers:

$$\delta V_{i,G}(r) = \frac{V_i(r) - V_G(r)}{V_G(r)} \tag{18}$$

where the subscript $i$ will indicate the $I$ or $Q$ samples. If the flux–limited samples do trace well the full galaxy dipole then we should find $\delta V(r) \simeq 0 \ \forall \ r$.

#### 4.1.1 Cluster (C) and 3D galaxy (G) dipole shapes

We find that the dipole shapes of clusters with varying $\langle d_C \rangle$ are extremely similar and out of the 106 LG–like observers only in 3 cases the corresponding dipole structures have opposite characteristics (decreasing vs. increasing amplitudes). The correlation coefficient analysis (see Table 2) gave $\langle c_{C_{38},C_{30}} \rangle \simeq 0.9 \pm 0.1$, while its median value is $\simeq 0.94$. As expected, the individual sample dipole amplitudes vary because of their different biasing parameters (see eq.9). Furthermore, for most of the 106 LG–like observers the full galaxy ($G$) dipole closely matches that of the cluster dipoles. The value of the corresponding correlation coefficient is $\langle c_{C_{38},G} \rangle \simeq 0.7 \pm 0.3$, with $\simeq 0.84$ for its median.

In Figure 2 we present a comparison between the $G$ (thick line), the $C_{38}$ (filled dots) and the $C_{30}$ (open dots) dipoles for 6 characteristic observers. Note that the cluster dipoles are scaled downwards by a factor of 2.5 and 2 for the $C_{38}$ and $C_{30}$ cases, respectively. Panels ($a$) to ($d$) show the most common case where the cluster dipole matches the shape of the underlying galaxy dipole rather well. Out of 106 observers only in 9 we do detect a strong disagreement in dipole morphology, convergence and direction. Two such examples are shown in panels (e) and (f) of Figure 2. We have verified that this discrepancy should be attributed to two facts:

- The $G$ dipole is dominated by a very nearby, although relatively low, density enhancement which is not represented in the $C_{38}$ clusters;
- The galaxy dipole shape is steeply increasing within $\sim 50$ $h^{-1}$ Mpc followed by a significant decrease, converging rather rapidly to a very low overall value.

Note that in the particular case shown in panel (f) the $C_{30}$ dipole shape is in good agreement with that of the galaxy one, although the same is not true for the $C_{38}$ dipole: this just implies that the fluctuations generating the galaxy dipole have passed the density threshold defining the $C_{30}$ cluster sample but not that of the $C_{38}$ sample.

### 4.1.2 Galaxy (G) and flux–limited (I and Q) dipole shapes

Regarding the flux–limited samples ($I$ and $Q$) we obtain a very good correlation of their dipole shapes with that of the full $G$ dipole. The correlation coefficients are always $\gtrsim 0.85$, but with a decreasing trend as a function of distance, which indicates that the $Q$ and $I$ samples miss some of the distant $G$ dipole contributions. We can identify 3 general cases, which are (in decreasing order of frequency):

(i) $\delta V_{Q,G} \lesssim \delta V_{I,G} < 0$, in $\sim 66\%$ of the observers. The $I$ and $Q$ samples underestimate the total $G$ dipole and in some cases even up to $\sim 30\%$. The dipole convergence depth is also underestimated, although in most cases the contributions to the dipole from large depths are apparent in both $I$ and $Q$ samples.

(ii) $\delta V_{I,G} \simeq \delta V_{Q,G} \approx 0$ (within 5%) in $\sim 26\%$ of the observers implying that in this case the $I$ and $Q$ samples trace accurately the $G$ dipole.

(iii) $\delta V_{I,G} \simeq \delta V_{Q,G} > 0$ in very few cases. Most such observers have a very low asymptotic $G$–dipole value ($\lesssim 200$ km s$^{-1}$) and therefore they also have a shot–noise dipole of the same order.

In Figure 3 we plot the $I$ dipole (dashed line), the $Q$ dipole (dotted line) and the full $G$ dipole (thick line) for the same observers as in Figure 2. In the inner panels we plot $\delta V_{G,I}$ (dashed line) and $\delta V_{G,Q}$ (dotted line). Note that the observer of panel (a) is representative of case (ii), those of panels (b) to (d) of case (i), that of panel (e) of case (iii) and finally the observer of panel (f) is an intermediate of cases (ii) and (iii).

In Table 2 we present an overview of the correlation coefficient analysis results. Deviations from the quoted mean values are $\sim 0.25$ when using all the 106 observers; by using the 95 observers of cases (i) and (ii) it lowers to $\sim 0.15$. Evidently, the high correlation coefficients indicate the strong correlation of dipole shapes, discussed previously.

Finally we plot, in Figure 4, the $\delta V$ values as a function of distance for all pairs of galaxy mass tracer

[tp]

**Table 2.** Correlations coefficients, $c_{ij}$, of dipole shapes for pairs of different samples. Results are averaged over all 106 observers.

|   | $G$ | $I$ | $Q$ | $C_{38}$ | $C_{30}$ |
|---|---|---|---|---|---|
| $G$ | – | 0.86 | 0.88 | 0.69 | 0.75 |
| $I$ | 0.86 | – | 0.89 | 0.56 | 0.63 |
| $Q$ | 0.88 | 0.89 | – | 0.74 | 0.76 |

samples. Errorbars correspond to $\pm 1\sigma$ estimates. A systematic underestimate, $\langle \delta V \rangle \sim 20\%$, of the true dipole by both the $I$ and $Q$ samples is evident.

### 4.1.3 Dipole directions and alignments

Another important indication of how well the underlying galaxy ($G$) dipole is traced by the dipoles of the different mass tracers is their respective dipole misalignment angle, $\Delta\theta$. If linear perturbation theory of eq.(3) is to be used to relate velocity and acceleration, then a necessary prerequisite is a good alignment of the velocity and acceleration (dipole) vectors. To this end, we calculate the dipole misalignment angles also as a function of distance.

In Figure 5 (panel a) we present the $\Delta\theta$ values as a function of distance for all pairs of mass tracers. As expected, we find that on average the distance at which the dipole direction converges to its final value coincides with of the dipole amplitude. At smaller depths, the misalignment angles fluctuate between $(55-70)^\circ$. Note, however, that in the true cluster case the CMB and cluster dipoles are well aligned already from within $\sim 50 - 60$ $h^{-1}$ Mpc, well before the final dipole amplitude convergence depth, which suggests that the LG motion is induced by a large–scale coherent anisotropy; see the discussion in Branchini & Plionis 1996 and references therein.

Between the cluster samples and the $G$, $I$, $Q$ samples we have $\langle \Delta\theta \rangle \sim 48^\circ (\pm 26^\circ)$, whereas for the two populations of clusters $\langle \Delta\theta_{C_{38}, C_{30}} \rangle \sim 23^\circ (\pm 20^\circ)$ with a median of $\sim 19^\circ$ which shows that the dipoles of the two cluster samples are consistent with each other in both shape and direction.

The large value of $\Delta\theta$ between clusters and galaxies (Fig. 5a) can be attributed in general to the sparse sampling of the underlying density field by the high peaks (cf. Tini Brunozzi et al. 1995) and in particular to the fact that the clusters, due to their large mean separation, will tend to miss the *local* (near to the observer) contributions to the galaxy dipole. In fact if we analyse only those observers that trace better the *local* density fluctuations, simply requiring that their cluster dipole acquires its first contribution from distances $\lesssim \langle d_c \rangle$, we find $\langle \Delta\theta \rangle \sim 32^\circ (\pm 14^\circ)$.

The relevant $\Delta\theta$ between different galaxy samples reveal a somewhat more complicated picture (see Figure 5b). The most significant feature is their systematic increase with radial distance which becomes more evident after the dipole convergence scale. We ascribe this to the systematic underestimate of the underlying galaxy density field due to dilute sampling, a phe-nomenon which is much stronger at large distances, thus producing large shot–noise effects and bigger deviations. However, different sparse samples define a different degree of shot–noise amplitude, which reflects inevitably in the way these samples follow the 3D density field. At $R = 180 \ h^{-1}$ Mpc, we find $\langle \Delta \theta_{G,I} \rangle \approx 17° (\pm 11°)$ and $\langle \Delta \theta_{G,Q} \rangle \approx 32° (\pm 36°)$, respectively. Note that the latter value is very close to the misalignment angle between the real QDOT and the CMB dipoles (RR90; Plionis et al. 1993). A further indication that the flux–limited samples underestimate the true $G$ dipole is also provided by the fact that the misalignment angle increases as a function of distance. For example, evaluating $\langle \Delta \theta_{G,I} \rangle$ and $\langle \Delta \theta_{G,Q} \rangle$ at 240 $h^{-1}$ Mpc we find $\sim 26° (\pm 19°)$ and $\sim 39° (\pm 36°)$, respectively.

### 4.1.4  General comments

According to above results, we can safely conclude that, in general, cluster samples (of Abell/ACO or APM type) and flux–limited galaxy samples (of IRAS/QDOT type) yield estimates of the dipole which are quite consistent with each other, and with the underlying dipole. Their amplitudes differ by a roughly constant biasing factor (as we will quantify below), although with a relatively large observer–to–observer scatter. However we have found a quite large average misalignment angle between the cluster and galaxy dipoles. Note that in the observed case (Plionis & Valdarnini 1991; Scaramella et al. 1991; Branchini & Plionis 1996) the Abell/ACO cluster dipole is very well aligned with that of the CMB ($< 25°$). We find only 22 such observers in our simulations, which constitute 21% of the whole sample, a rather small but non negligible fraction. For these observers we then find $\langle c_{C_{3s},G} \rangle \simeq 0.9$ and $\langle c_{C_{3s},I} \rangle \simeq 0.8$, i.e. correlation coefficients between the cluster and galaxy dipole shapes which are much larger than in the general observer case.

In the galaxy flux–limited cases (either the $I$ or $Q$ samples) we found very good agreement of their respective dipole shapes and alignments, although there is also strong evidence for an underestimation of the underlying ($G$) dipole contributions from large–scales.

We summarize our conclusions on these mass tracers as follows.

• The increase of $\langle \delta V \rangle$ and $\langle \Delta \theta \rangle$ as a function of distance for the $[G, I]$ and $[G, Q]$ cases indicates that IRAS–like flux–limited samples underestimate the dipole contributions from large depths.

• The value of $\langle \delta V_{G,I} \rangle \simeq \langle \delta V_{G,Q} \rangle \approx -0.2$ indicates that both IRAS and QDOT samples will miss roughly the same fractional amount of the full dipole, independently of the 1–in–6 or 6–in–6 sampling.

• The value of $\langle \Delta \theta \rangle$ is, however, larger for the 1–in–6 than the full 6–in–6 flux–limited samples by, on average, $\sim 15°$.

### 4.2  A Statistical Biasing Scheme

Various authors have used so far different methods to estimate the bias parameter. Correlation function analysis (cf. Lahav, Nemiroff & Piran 1990), power spectrum techniques (cf. Jing & Valdarnini 1993; Peacock & Dodds 1994) and ratio of count–in–cell variances at different scales (cf. Borgani et al. 1995). Here we describe our method to estimate $b$ which is similar to that of Plionis (1995) who estimated the cluster–IRAS relative bias factor using the Abell/ACO and QDOT dipoles in the range $10 \lesssim r \lesssim 100 \ h^{-1}$ Mpc: he found, after correcting for redshift space distortions, that $b_{C,I} \approx 3.5$.

As discussed in Section 2.1, assuming linear perturbation theory, we can in principle relate the simulated observer peculiar velocities, $u$, to those predicted from their dipole, $v$, as measured using eq.(9), for a particular class of mass tracers. Therefore, comparing observer–by–observer the $v$ values for different mass tracers one can estimate their relative biasing. Note that since in the simulation we identify galaxies by a random sub-sampling of the dark matter (DM) particles, we have by definition that $b_G = b_I = b_Q = 1$ for the $G, I, Q$ biasing factors: any departure from unity value is merely to be ascribed to sampling fluctuations. Therefore, in the remaining analysis we will consider only the relative bias between clusters and galaxies. Since $\Omega_\circ = 1$ in our simulation, we get from eq.(9) that

$$u(r) = v_G(r) = b_C^{-1} v_C(r) \ , \tag{19}$$

where $b_C$ is the cluster biasing factor. Therefore, we can obtain an estimate of $b_C$ by fitting eq.(19) over the distance range where the dipole is estimated, by minimizing the quantity

$$\chi^2(b, \gamma) = \sum_{i=1}^{N} \left( \frac{v_C - b_C \ v_G - \gamma}{\sigma_i} \right)^2 \ . \tag{20}$$

Here the sum is over the $N$ radial bins used to estimate the cumulative mass tracer dipoles and $\gamma$ represents a 'zero–point' uncertainty which is bound to exist due to the fact the mean separation is much larger for clusters than for galaxies and thus the cluster dipole will always miss the very local contributions, if such do exist (see Plionis 1995 for an application to real data). It is also evident that we cannot obtain an estimate of the goodness of fit since the bins are not independent.

Note that since eqs.(3) and (9) are of doubtful applicability when a large dipole misalignment is measured, we use in this analysis only those observers for which $\Delta \theta_{CG} \leq 25°$. We will then average the results over the ensemble of such observers. In Tables 3 and 4 we list the mean values and the $1\sigma$ scatter for the $b$ and $\gamma$ parameters, respectively, as obtained from the $\chi^2$–minimization method (in which we use $\sigma_i = 1$). The average formal fitting error in $b$ and $\gamma$ was found to be $\sim 0.04$ and $\sim 200$ km s$^{-1}$, respectively.

In order to check the reliability of using the dipole amplitudes to estimate the cluster biasing factor, we computed it also by the count–in–cell method (see also Tini Brunozzi et al. 1995). We randomly displaced 5000 spheres within the simulation box, with radii ranging from $50 \ h^{-1}$ Mpc to $120 \ h^{-1}$ Mpc and estimated the r.m.s. value, $\sigma_{C_{3s}}(R)$ of the cluster count within such spheres. Therefore, the cluster biasing factor can be statistically defined as

[tp]

**Table 3.** Relative biases between clusters and galaxies.

| SAMPLES | $G$ | $I$ | $Q$ |
|---------|-----|-----|-----|
| $C_{38}$ | 4.1± 1.1 | 4.2± 1.6 | 4.8± 1.6 |
| $C_{30}$ | 3.3± 0.6 | 3.4± 0.9 | 3.8± 0.8 |

[tp]

**Table 4.** Relative $\gamma$ parameters (in km s$^{-1}$) between clusters and galaxies.

| SAMPLES | $G$ | $I$ | $Q$ |
|---------|-----|-----|-----|
| $C_{38}$ | -520±360 | -290±500 | -460±315 |
| $C_{30}$ | -220±390 | -50±440 | -170±380 |

$$b_{C_{38}} = \frac{\sigma_{C_{38}}}{\sigma_{DM}}. \qquad (21)$$

Here $\sigma_{DM}$ is the r.m.s. fluctuation for the DM distribution, which we computed from the power spectrum $P(k)$ according to

$$\sigma_{DM}(R) = \left[\frac{1}{2\pi^2}\int_0^\infty dk\, k^2\, P(k)\, W^2(kR)\right]^{1/2}, \qquad (22)$$

with $W(kR)$ the Fourier transform of the top–hat sphere, $W(x) = 3(\sin kR - kR\cos kR)/(kR)^3$. We find that the cluster biasing factor is constant to a good accuracy over the whole scale range, having a value $b_{C_{38}} = 3.9 \pm 0.2$, in nice agreement with the result reported in Table 3.

A further point in this analysis concerns the large values of $\gamma$ (Table 4) which one would have naively expected to be rather close to zero. However as we discussed earlier, these $\gamma$ values just reflect the cluster dipole 'zero–point' uncertainty which is due to the fact the cluster distribution cannot in general accurately trace the galaxy distribution near the observer, therefore missing local dipole contributions. A similar situation appears in the real Abell/ACO cluster case, where the cluster dipole cannot trace the contribution to the LG motion of the Virgo cluster; see Plionis 1995 for an attempt to correct for such effects.

However, the method we are using is only a statistical measure of the relative biases, based on various assumptions posed by the corresponding numerical experiments. Although the various uncertainties (luminosity function variations, large shot–noise errors due to sparse sampling, etc.) may cause a possible poor estimate of the relative biases, when added together, the fact that we actually manage to recover the mean bias factors so close to those measured by other methods, should be considered a success for this approach.

## 5  CONCLUSIONS

We have used numerical simulations of a Cold+Hot dark matter model, based on the Truncated Zel'dovich Approximation, to simulate the distribution of rich clusters, of IRAS galaxies and of flux–limited subsamples of IRAS galaxies in order to examine how well the different mass tracers reflect the underlying "galaxy" dipole. We address also the effects of the QDOT sparse sampling strategy to the determination of the underlying dipole structure.

Two cluster samples were generated; the first with a mean intercluster separation of $\langle d_C \rangle = 38\ h^{-1}$ Mpc (in order to mimic Abell/ACO sample) and the second with $\langle d_C \rangle = 30\ h^{-1}$ Mpc (resembling the APM sample). Galaxies are identified from a random subsampling of the whole DM particle distribution, having, however, the same average density and radial selection functions as the real IRAS and QDOT samples. We have allowed ourselves such a simplification of the galaxy identification procedure since we are interested only in analysing differences between the dipole structures of different mass–tracer distributions, rather than in the details of the distribution of "true" simulated galaxies in a specific DM model. We chose observers measuring a cluster dipole similar to the Abell/ACO one; then we calculated for each of them the dipoles for all the catalogues of mass tracers. Care has also been taken to estimate and correct for the relative shot–noise errors generated by the sparse sampling.

Using a a statistical correlation coefficient analysis we found that in most cases the cluster dipole shape reflects a similar 3D galaxy dipole shape while the alignment of the two vectors is fair but not extremely good, $\langle\Delta\theta_{C_{38},G}\rangle \sim 48°$. Such large misalignment angles should be attributed to the fact that clusters are high peaks of the underlying density field and therefore, due to their large mean separation, they inevitably tend to miss the local (i.e. near to the observer) gravitational contributions, which in most cases dominate the galaxy dipole. Consistently, if we restrict the analysis to those observers for which the cluster distribution traces fairly well the local galaxy dipole contributions we systematically obtain a better alignment, $\langle\Delta\theta_{C_{38},G}\rangle \sim 32°$.

Comparing the dipoles obtained from the flux–limited galaxy samples and from the whole galaxy distribution, we find a close match of their shapes for both IRAS– and QDOT–like samples. However, both the flux–limited samples miss a fraction of the total dipole amplitude (on average $\sim 15\% - 20\%$). The average misalignment angle between the dipoles of the whole distribution and of the IRAS–like sample is $\langle\Delta\theta\rangle \simeq 17°$; for the QDOT–like case it is $\langle\Delta\theta\rangle \simeq 32°$, similar to the observed value between QDOT and CMB dipoles (RR90; Plionis et al. 1993). It should be noted that the quoted misalignment angles between the galaxy samples are not shot–noise–free; such corrections have only been applied to their dipole amplitude estimates.

As a main conclusion of this paper, we predict that the dipole of the full 6–in–6 QDOT sample will have about the same amplitude as the original 1–in–6 sample, or in other words it will underestimate the true dipole by the same amount as the 1–in–6 sample (by $\sim 15\% - 20\%$). However, the alignment of the 6–in–6 dipole with that of the CMB should become better and within $\sim 10° - 17°$.

Finally, we have used a statistical method, based on linear perturbation theory and on the linear biasing assumption given by eq.(1), to estimate the relative bias factors of the different mass tracers. Our results agree well with those based on the count–in–cell statistics, in-dicating that our method is robust, although it produces quite a large observer–to–observer scatter in the derived biasing parameters.

## ACKNOWLEDGEMENTS

VK receives a PPARC research studentship. PC is a PPARC Advanced Research Fellow. MP has been supported by an EEC *Human Capital and Mobility* fellowship. LM thanks Italian MURST for partial financial support. We thank Dimitra Rigopoulou, Enzo Branchini and Paolo Tini Brunozzi for useful discussions during the course of this work. We also acknowledge financial support from the European Union HCM programme under contract ERBCHRX–CT93–0129.

**FIGURE CAPTIONS**

**Figure 1.** Number density of the mock IRAS galaxies as a function of distance. The superimposed line corresponds to that expected in a perfectly homogeneous universe.

**Figure 2.** Cumulative dipole amplitudes for 6 different observers. The $C_{38}$ and $C_{30}$ cluster dipoles are represented by the filled and open dots, respectively. The thick line corresponds to the full underline galaxy ($G$) dipole.

**Figure 3.** The $G$ (thick line), $I$ (dotted line) and $Q$ (dashed line) galaxy dipoles for the same observers as in Figure 2. The inner panels show their velocity fluctuations $\delta V$.

**Figure 4.** Mean velocity fluctuations, $\langle \delta V \rangle$, as a function of radial depth. Errorbars are $1\sigma$ scatter (shown only for a few points for clarity).

**Figure 5.** Mean misalignment angles, $\langle \Delta \theta \rangle$, between $C_{38}$ clusters and galaxies (panel $a$), and between the galactic samples, $G$ vs. $I$ and $G$ vs. $I$ (panel $b$), are shown as a function of depth. The squares in panel ($a$) represent $\langle \Delta \theta \rangle$ between the $C_{38}$ and $C_{30}$ clusters. Errorbars, plotted only for one set of points, are similar in the other cases.